\documentclass[a4paper,noarxiv,onecolumn]{quantumarticle}
\usepackage[utf8]{inputenc}
\usepackage[english]{babel}
\usepackage[T1]{fontenc}
\usepackage{amsmath}
\usepackage{hyperref}
\usepackage{graphicx} %
\usepackage{subfigure} %
\usepackage{lipsum}
\usepackage{mdframed}
\usepackage{booktabs}
\usepackage{longtable}

\title{Deterministic entanglement swapping of W states}
\author{Sajede Harraz}
\email{sajede@ustc.edu.cn}
\affiliation{University of Science and Technology of China, Hefei, China}

\author{Wang Yueyan}
\email{wyy\_00@mail.ustc.edu.cn }
\affiliation{University of Science and Technology of China, Hefei, China}

\author{Shuang Cong}
\email{scong@ustc.edu.cn}
\affiliation{University of Science and Technology of China, Hefei, China}

\begin{document}

\maketitle

\begin{abstract}
In this paper, we propose a deterministic entanglement swapping protocol for generating a shared three-qubit W state between two remote parties. Our method offers a reliable alternative to existing probabilistic protocols for W state entanglement swapping, which is crucial for various quantum information processing tasks. We present a detailed quantum circuit design, implemented using the Qiskit simulator, that outlines the preparation of W states and the execution of joint measurements required for the entanglement swapping process. Furthermore, we analyze the effects of imperfect operations and noisy communication channels on the fidelity of the resulting shared W state. To address these challenges, we introduce a weak measurement-based purification method that enhances fidelity in the presence of amplitude damping. Through mathematical analysis and Qiskit simulations, we demonstrate the effectiveness of our proposed protocol, offering a practical solution for high-fidelity W state generation in real-world quantum communication scenarios.
\end{abstract}

\section{Introduction}

Creating reliable connections between distant users in quantum communications faces a major challenge due to the weakening of entanglement between particles as the distance increases. Entanglement swapping offers a solution by creating entanglement between particles that have never interacted before, allowing for more dependable long-distance quantum communication. As a key part of quantum repeaters and relays \cite{lab1, lab1a, lab2}, entanglement swapping plays a significant role in developing quantum communication technologies. Usually involving three separated users, two users share a pair of entangled particles with the third user. When the third user performs a joint measurement on these particles, the other two users' particles become entangled, even though they were not initially connected. This process highlights the importance of entanglement swapping in quantum relay systems \cite{lab3, lab4}. Integrating entanglement swapping with quantum purification \cite{lab5, lab6, lab6a} and quantum memory \cite{lab7, lab8} forms a complete quantum repeater. This combination enables practical quantum communication by overcoming noise and imperfections, ensuring more efficient and reliable quantum information exchange.

Despite extensive research on entanglement swapping for Bell and GHZ states, it is essential to recognize the unique significance of W-states in quantum communication. W-states possess a distinctive resilience against particle loss; when one particle is lost, entanglement persists between the remaining two particles—a characteristic not exhibited by GHZ states. This attribute highlights the importance of W-states in quantum communication, prompting the proposal of several approaches for their generation \cite{lab9, lab10, lab11}. However, there remains a research gap concerning entanglement swapping protocols specifically designed for distributing W-states among quantum communication parties. A recent development in this area is the introduction of a probabilistic swapping protocol for W-states within a two-dimensional triangular quantum network, achieving an overall protocol success probability of $\frac{2}{3}$ \cite{lab111}.

The generation of entangled quantum states in experimental settings may also suffer from non-ideal outcomes and potential degradation when transmitted through communication channels \cite{lab112}. Amplitude damping is a critical form of decoherence affecting various platforms \cite{11a, 11aa}. For instance, in trapped-ion optical qubits, amplitude damping arises from photon scattering in a metastable level. Moreover, residual photon scattering in entangling gates utilizing two-photon processes through far-detuned dipole-allowed transitions also results in amplitude damping \cite{11b, 11c, 11d}. Notably, the operator elements of amplitude damping cannot be represented using scaled Pauli matrices, complicating the purification process.

Enhancing the degree of entanglement in shared entangled states requires additional operations, for which several strategies have been proposed. Entanglement distillation \cite{lab12a, lab12b, lab12c} and quantum error correction codes \cite{lab12d, lab12e, lab12f} are among these approaches. However, these methods have their own set of challenges, as they necessitate a significant number of redundant qubits or a substantial quantity of identically prepared entangled qubits, rendering them resource-intensive and challenging to implement in practice. In contrast, weak measurement-based protection schemes provide an alternative approach to mitigating the harmful effects of decoherence on entangled states \cite{lab121, lab122, lab12h}. Compared to entanglement distillation, these schemes only require a single copy of the entangled state, significantly reducing the resource demands associated with the distillation process. Consequently, weak measurement-based protection schemes present a more practical and resource-efficient solution for improving the entanglement of shared states.

In this paper, we present a deterministic entanglement swapping protocol tailored specifically for W-states. By harnessing a particular class of W states, our protocol guarantees deterministic entanglement swapping, offering a more dependable and efficient solution for quantum communication networks compared to probabilistic methods. We develop a custom Qiskit circuit that facilitates the preparation of these targeted W states and executes the joint measurements required for the entanglement swapping process\footnote{\textcolor{black}{Matlab and Qiskit codes for regenerating the results of this article are available from "https://github.com/Sjd-Hz/W-state-entanglement-swapping".}}. Our proposed protocol's effectiveness is substantiated through mathematical analysis and Qiskit simulations, ensuring its reliability and relevance in real-world applications. Moreover, we explore the effects of imperfect operations and noisy communication channels on the fidelity of the resulting shared W state. To address these challenges, we introduce a purification method based on weak measurements, aiming to enhance the fidelity of the shared entangled state under amplitude damping conditions. This approach effectively mitigates the detrimental impact of amplitude damping and improves the fidelity of the shared entangled state, as demonstrated by mathematical analysis and numerical and Qiskit simulation results. Our findings contribute to the advancement of quantum communication technologies, paving the way for more robust and efficient quantum networks.

The remainder of the paper is structured as follows: Section \ref{Sec1} offers a comprehensive description of the proposed entanglement swapping protocol for W states, including the custom-designed Qiskit circuit that facilitates the preparation of the targeted W states and the execution of joint measurements required for the entanglement swapping process. Section \ref{Sec2} examines the impact of imperfect operations and noisy communication channels on the fidelity of the resulting shared W state. Additionally, it provides a thorough analysis of the weak measurement-based purification method, which aims to address the challenges posed by amplitude damping and enhance the fidelity of the shared entangled state. Finally, Section \ref{Sec3} presents the conclusion, summarizing the key findings and contributions of the paper.

\section{Deterministic entanglement swapping for W states}\label{Sec1}

In this section, we will discuss the specifics of our proposed entanglement swapping protocol for W states. We should mention that the prototype W states $|{{\text{W}}}{\rangle _{123}} = \frac{1}{{\sqrt{3} }}{(|100\rangle _{123}}+ |010{\rangle _{123}} + |001{\rangle _{123}})$, as examined in \cite{lab111}, results in a probabilistic swapping protocol. In contrast, our proposed protocol utilizes a distinct class of W states, denoted as $|{{\text{W}_n}}{\rangle}$, which falls within the category of W states suitable for serving as an entanglement resource. This class of states is defined as 
\begin{equation}
\label{Eq6}
\begin{aligned}
|{{\text{W}_n}}{\rangle} =& \frac{1}{{\sqrt {2 + 2n} }}{(|100\rangle}+ \sqrt n {e^{i\gamma }}|010{\rangle} + \sqrt {n + 1} {e^{i\delta }}|001{\rangle }),
\end{aligned}
\end{equation}
where $n \in {\bf{R}}^+$, and $ \gamma, \delta \in {\bf{R}}$ are global phases. Considering the simplified case where $n=1$ and assuming zero phases, the W state can be represented as \cite{lab12}
\begin{equation}
\label{Eqw}
|{{\text{W}}}\rangle = \frac{1}{{2 }}{(|100\rangle}+ |010{\rangle } + \sqrt {2} |001{\rangle }).
\end{equation}

By consulting the criteria provided in \cite{lab13}, we can confirm that this state belongs to the category of W states.

\begin{figure}[!htpb]
    \centering
    \includegraphics[width=\linewidth]{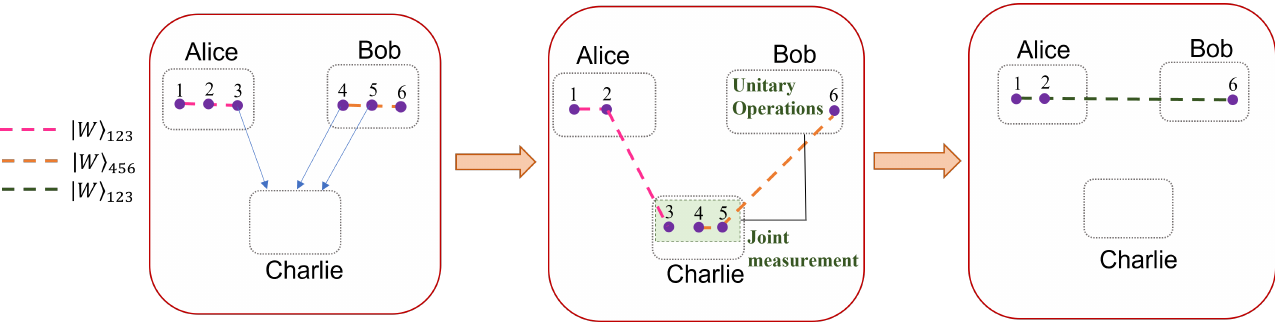}
    \caption{A schematic diagram of the entanglement swapping process.}
    \label{schem}
\end{figure}

The entanglement swapping procedure involves three parties: Alice, Bob, and Charlie. The ultimate objective is to establish a shared entangled W state between Alice and Bob. Initially, Alice and Bob each share an entangled state with Charlie. Charlie performs a joint measurement on the qubits from both Alice's and Bob's entangled states and communicates the results to Bob. Subsequently, Bob applies the corresponding unitary rotation operations on his qubit. As a result, an entangled shared state between Alice and Bob is induced. This process, known as entanglement swapping, allows the creation of an entangled connection between Alice and Bob without direct interaction between them. The schematic diagram of the entanglement swapping process is presented in Fig. \ref{schem}.

The details of our proposed entanglement swapping process are as follows:

Alice and Bob independently prepare W states in Eq. \eqref{Eqw} at their respective locations. The combined state is given by:
\begin{widetext}
\begin{equation}\label{eqc}
\begin{aligned}
&|{\psi}{\rangle _{\text{combined}}} =|{{\text{W}}}{\rangle _{123}} \otimes|{{\text{W}}}{\rangle _{456}}\\
&=\frac{1}{{2 }}{(|100\rangle _{123}}+ |010{\rangle _{123}} + \sqrt {2} |001{\rangle _{123}})\otimes \frac{1}{{2 }}{(|100\rangle _{456}}+ |010{\rangle _{456}} + \sqrt {2} |001{\rangle _{456}})\\
&=\frac{1}{4}{(|100\rangle_{123}|100\rangle _{456}}+|100\rangle_{123}|010\rangle _{456}+\sqrt{2}|100\rangle_{123}|001\rangle _{456}\\
&+|010\rangle_{123}|100\rangle _{456}+|010\rangle_{123}|010\rangle _{456}+\sqrt{2}|010\rangle_{123}|001\rangle _{456}\\
&+\sqrt{2}|001\rangle_{123}|100\rangle _{456}+\sqrt{2}|001\rangle_{123}|010\rangle _{456}+2|001\rangle_{123}|001\rangle _{456}),
\end{aligned}
\end{equation}
\end{widetext}
where qubits 1, 2, and 3 belong to Alice, and qubits 4, 5, and 6 belong to Bob. To initiate the entanglement swapping process, Alice sends her last qubit (qubit 3) and Bob sends his first two qubits (qubits 4, 5) to Charlie. Hence, by rearranging the qubits in Eq. \eqref{eqc}, the combined state shared between Alice, Bob, and Charlie can be represented as:
\begin{widetext}
  \begin{equation}
  \begin{aligned}
&|{\psi}{\rangle _{\text{combined}}}=\frac{1}{4}(|010\rangle _{345}|100\rangle_{126}+|001\rangle _{345}|100\rangle_{126}+\sqrt{2}|000\rangle _{345}|101\rangle_{126}\\
&+|010\rangle _{345}|010\rangle_{126}+|001\rangle _{345}|010\rangle_{126}+\sqrt{2}|000\rangle _{345}|011\rangle_{126}\\
&+\sqrt{2}|110\rangle _{345}|000\rangle_{126}+\sqrt{2}|101\rangle _{345}|000\rangle_{126}+2|100\rangle _{345}|001\rangle_{126})\\
&=\frac{1}{2}[|{\eta ^+}\rangle_{345}\otimes|{W}\rangle_{126} +|{\eta^-}\rangle_{345} \otimes (I\otimes I \otimes \sigma_z)  |{W}\rangle_{126}+\\
&|{\xi^+}\rangle_{345} \otimes (I\otimes I \otimes \sigma_x)  |{W}\rangle_{126}+|{\xi^-}\rangle_{345} \otimes (I\otimes I \otimes \sigma_x\sigma_z)  |{W}\rangle_{126}],
  \end{aligned}
  \end{equation}
\end{widetext}

where $|{\eta ^{\pm}}\rangle, |{\xi ^{\pm}}\rangle$ are a set of orthogonal states in the W state category given as
\begin{equation}\label{orth}
\begin{aligned}
|{\eta ^{\pm}}\rangle &= \frac{1}{2}(|010\rangle + |001\rangle  \pm \sqrt 2 |100\rangle ),\\
|{\xi^{\pm}}\rangle  &= \frac{1}{2}(|110\rangle  + |101\rangle  \pm \sqrt 2 |000\rangle ).
\end{aligned}
\end{equation}

Charlie performs a von Neumann measurement on the combined system of three particles 345 using the basis comprising the states $\{|{\eta ^{\pm}}\rangle, |{\xi ^{\pm}}\rangle\}$. Upon obtaining the measurement result, he encodes it into two classical bits and transmits them to Bob. Utilizing the received information, Bob applies a specific unitary transformation from the set $\{I,\sigma_z,\sigma_x,\sigma_x\sigma_z\}$ to the entangled state shared between him and Alice. By applying this operation, the shared state is effectively transformed into the W state represented by Eq. \eqref{Eqw}. Notably, this transformation occurs in a deterministic manner, ensuring a reliable and consistent outcome in the entanglement swapping process.

Here, we note that by considering the class of W states in Eq. \eqref{Eq6}, the general orthogonal states would be written as
\begin{widetext}
\begin{equation}
\begin{aligned}
|{\eta ^{\pm}_n}\rangle &= \frac{1}{\sqrt {2 + 2n}}(|010\rangle + \sqrt n {e^{i\gamma }}|001\rangle  \pm \sqrt {n + 1} {e^{i\delta }}|100\rangle ),\\
|{\xi^{\pm}_n}\rangle  &= \frac{1}{\sqrt {2 + 2n}}(|110\rangle  + \sqrt n {e^{i\gamma }}|101\rangle  \pm \sqrt {n + 1} {e^{i\delta }} |000\rangle ).
\end{aligned}
\end{equation}
\end{widetext}
\subsection{Qiskit implementation}

In this section, we describe the quantum circuit we have developed for the preparation of the W state and the implementation of the joint measurements required for the entanglement swapping process.

\begin{figure}[!htpb]
    \centering
    \includegraphics[width=0.8\linewidth]{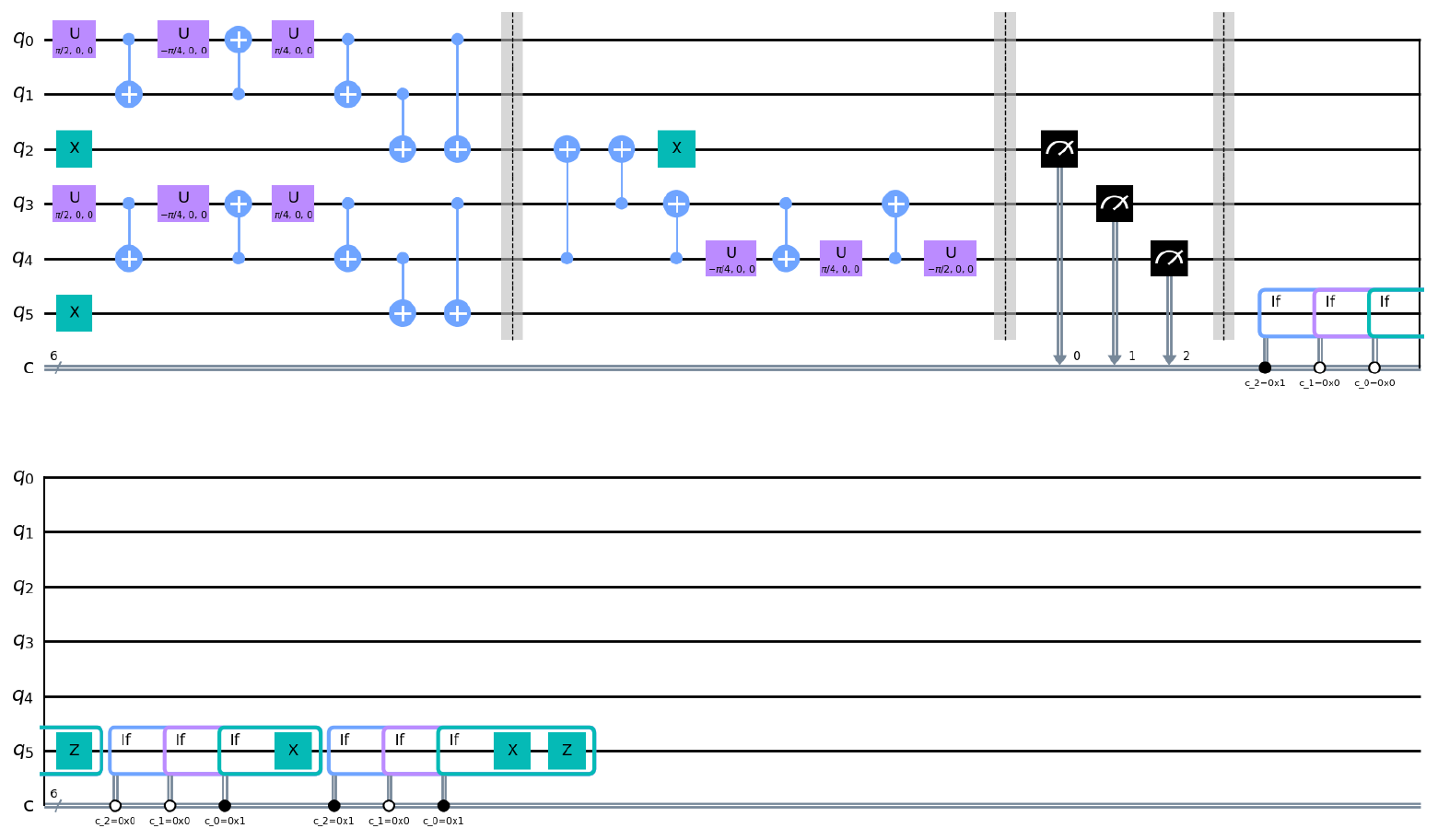}
    \caption{The Qiskit circuit for implementing the proposed W-state entanglement swapping.}
    \label{Scheme}
\end{figure}

In Fig. \ref{Scheme}, we present a Qiskit circuit designed for our proposed 3-qubit W-state entanglement swapping. Qubits $q_0$ and $q_1$ are located at Alice's site, $q_2$, $q_3$, and $q_4$ at Charlie's site, and $q_5$ at Bob's site. Vertical grey barriers are used to distinguish different phases of the circuit. Prior to the first barrier, the W-states are initialized according to Eq. \eqref{Eqw}. Following the first barrier, designed operations are applied to transform the orthogonal basis given in Eq. \eqref{orth} to the standard computational basis ($z$-basis). After the second barrier, Charlie measures the qubits at his location and informs Bob of the measurement results. The corresponding $\sigma_x$ and $\sigma_z$ operations are then applied at Bob's location after the third barrier.

\begin{figure}[!htpb]
    \centering
    \includegraphics[width=0.4\linewidth]{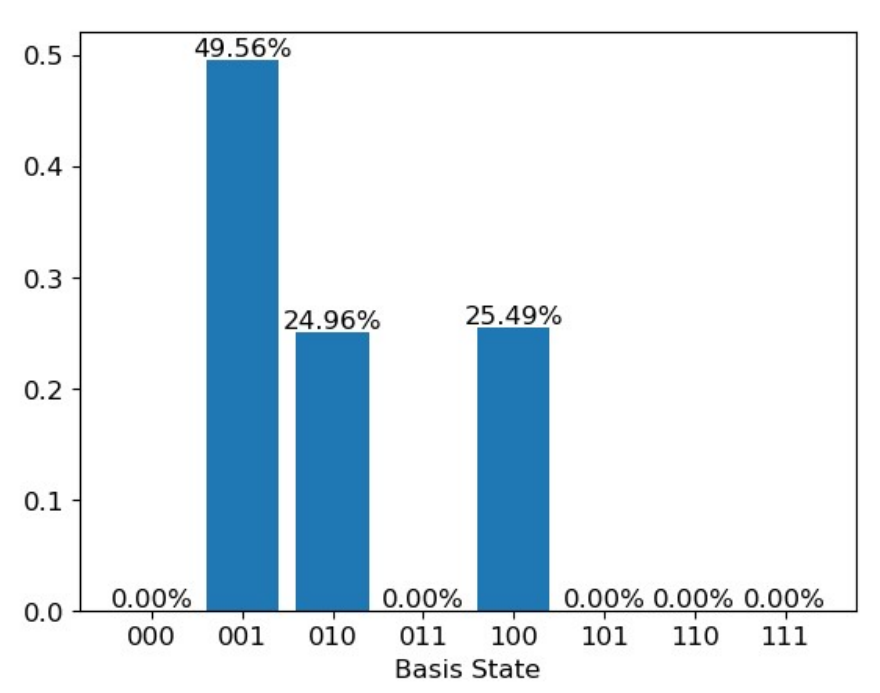}
    \caption{Average probabilities of basis states of the final shared entangled state between Alice and Bob.}
    \label{rho}
\end{figure}

By executing the Qiskit circuit on the AerSimulator and tracing out the qubits at Charlie's location, we selectively analyze states corresponding to each of Charlie's measurement outcomes. The average probability of basis states resulting from this process is depicted in Fig. \ref{rho}. It is evident that the shared entangled state aligns precisely with the W state in Eq. \eqref{Eqw}.

\section{Considering imperfect operations and noisy communication channels}\label{Sec2}

In the previous section, our analysis has been based on the assumption of ideal conditions, in which the communication channels between Alice and Bob were noise-free, and all quantum gate operations were performed perfectly. However, in practical scenarios, noise is invariably present, and the shared entanglement often experiences significant degradation due to various decoherence mechanisms that affect real implementations. 

\subsection{Imperfect operations}

This subsection evaluates the practicality and effectiveness of our proposed W-state entanglement swapping protocol under real-world conditions involving imperfect operations. To maintain a focused analysis, we do not consider memory errors and single-qubit gate errors, as these factors generally exhibit lower error rates. Our primary focus is on the  effects of imperfect CNOT operations, as two-qubit operations are more susceptible to errors in experimental setups \cite{14,15,16}. We assume a scenario where each CNOT gate has a probability of $y_2$ for correct execution, and a complementary probability of $1-y_2$ for complete depolarization of the affected qubits $i$ and $j$. In this scenario, the resulting state of the density matrix, when applied to the input state $\rho_{in}$, can be mathematically represented as 
\begin{equation}\label{CNot_er}
\begin{aligned}
    \rho_{out}= & y_2 \ \rm CNOT \ \rho_{in} \ \rm CNOT^{T}+(1- y_2)Tr_{i,j}(\rho_{in})\otimes\frac{I_{i,j}}{4},
\end{aligned}
\end{equation}
where $\text{Tr}_{i,j}$ represents a partial trace over the affected qubits and $I_{i,j}$ is the identity operator associated with qubits $i$ and $j$ \cite{17,18}.

Recognizing the importance of measurement accuracy for the proposed entanglement swapping protocol, we must consider the impact of imperfect measurements on the success of the process. Specifically, the measurement of qubit $i$ has a probability $\eta$ of resulting in an accurate projection and measurement, and a complementary probability $1-\eta$ of producing an incorrect outcome, leading to a flipped qubit in the measurement basis. This can be mathematically illustrated using the example of an imperfect measurement on $|0\rangle$:
\begin{equation}\label{measur_error}
    \rho_{out}=\eta|0\rangle\langle0|\rho_{in}|0\rangle\langle0|+(1-\eta)|1\rangle\langle1|\rho_{in}|1\rangle\langle1|.
\end{equation}

\begin{figure}[!htpb]
    \centering
    \includegraphics[width=0.4\linewidth]{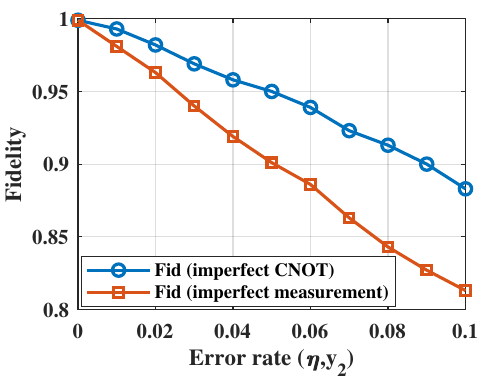}
    \caption{Fidelity of the final shared entangled state between Alice and Bob by considering imperfect CNOT and measurement operations.}
    \label{imperfect}
\end{figure}

Fig. \ref{imperfect} depicts the fidelity of the final shared W state between Alice and Bob under a realistic scenario where depolarizing error noise is introduced specifically for CNOT operations and readout error is applied to measurements on the qubits at Charlie's location, using the Qiskit Aer simulator.

As shown in Fig. \ref{imperfect}, imperfect measurement has a significantly greater impact on the fidelity of the final shared W state between Alice and Bob. This is due to the fact that unitary operations are applied based on the measurement results. Consequently, with imperfect measurements, unsuitable unitary operations may be employed, causing a more substantial decrease in the fidelity of the final shared entangled state compared to imperfect CNOT operations.

Here we note that several strategies have been developed to enhance the fidelity of the final shared entangled state, collectively referred to as entanglement purification methods \cite{lab111}. These methods involve utilizing multiple copies of the damped entangled state and employing local operations and classical communication to extract fewer entangled states with improved fidelity \cite{lab6}. In the following section, we will examine a specific type of decoherence known as amplitude damping, which presents a greater challenge as its operator elements cannot be expressed using scaled Pauli matrices.

\subsection{Weak measurement-based purification method for amplitude damping channels}

In this section, we focus on amplitude damping and propose a purification method based on weak measurements. This method seeks to provide an efficient and practical solution, overcoming the constraints associated with resource-intensive techniques traditionally used to manage decoherence-related challenges. We assume that Alice and Bob transmit their qubits to Charlie via amplitude damping channels (ADC), taking into account the realistic scenario of quantum information transmission. The amplitude damping is defined by the well-known Kraus operators as \cite{11a,11aa}:
\begin{equation}\label{eqer}
    e_0 = \begin{bmatrix} 1&0\\0&\sqrt{1-r} \end{bmatrix} \quad     e_1 = \begin{bmatrix} 0&\sqrt{r}\\0&0 \end{bmatrix},
\end{equation}
where \(r\in[0,1]\) is the decaying rate of the excited state with \(r=1-e^{-\Gamma t}\) and  \(\Gamma\) is the energy relaxation rate and \(t\) is the evolving time. This is an asymmetric channel because the qubit states $\left| {{1}} \right\rangle$ are transformed to $\left| {{0}} \right\rangle$ with probability $r$, while the $\left| {{0}} \right\rangle$ state never transforms to  $\left| {{1}} \right\rangle$, regardless of the value of $r$. The decaying rate of the amplitude damping is related to the natural lifetime of the qubit \cite{26a}.

Now let us analyze the impact of the ADC on the shared entangled state between Alice and Charlie, as well as Bob and Charlie. In accordance with the entanglement swapping procedure for the W state, Alice transmits her last qubit (qubits 3) through ADC, resulting in the following shared entangled state between Alice and Charlie:
\begin{equation}
    \begin{aligned}
        \rho_{123}^{\text{AD}} &= \sum_{i=0}^{1} (I \otimes I \otimes e_i)\left(|{W}\rangle_{123}\langle{W}|_{123}\right)(I \otimes I \otimes e_i)^{\dagger}\\
        &=|{W}\rangle_{123}^{\text{AD}} \langle{W}|_{123}^{\text{AD}} + \frac{r}{2} |{000}\rangle\langle{000}|,
    \end{aligned}
\end{equation}
where \(|{W}\rangle_{123}^{\text{AD}}=\frac{1}{2} \left( |{100}\rangle + |{010}\rangle + \sqrt{2(1-r)} |{001}\rangle \right)\).

Similarly, Bob sends his first two qubits (qubits 4, 5) through the ADC to Charlie, which modifies the shared entangled state between Bob and Charlie as follows:
\begin{equation}
    \begin{aligned}
        \rho_{456}^{\text{AD}} &= \sum_{i,j=0}^{1} (e_i \otimes e_j \otimes I) \left(|{W}\rangle_{456}\langle{W}|_{456} \right)  (e_i \otimes e_j \otimes I)^\dagger\\
        &=  |{W}\rangle_{456}^{\text{AD}} \langle{W}|_{456}^{\text{AD}}+  \frac{r}{2} |{000}\rangle \langle{000}|,
    \end{aligned}
\end{equation}
where $|{W}\rangle_{456}^{\text{AD}} = \frac{1}{2} \left( \sqrt{1-r} |{100}\rangle + \sqrt{1-r} |{010}\rangle + \sqrt{2} |{001}\rangle \right)$.

After sending the corresponding qubits through ADC to Charlie, the combined state shared between Alice, Bob, and Charlie is
\begin{equation}
\begin{aligned}   
    {\rho}^{\text{AD}}_{\text{combined}}&=\rho_{123}^{\text{ADC}}\otimes\rho_{456}^{\text{ADC}}.
\end{aligned}
\end{equation}

At this step, Charlie performs the joint measurements given in Eq. \eqref{orth} and sends the results to Bob, who then applies the corresponding unitary operations. 

It is important to mention that the fidelity of the final entangled state shared between Alice and Bob, considering the effects of ADC and the measurement outcome $|\eta^{\pm}\rangle$, can be enhanced by utilizing weak measurements. However, our proposed weak measurement-based purification method does not improve the fidelity of the states corresponding to the measurement outcome $|\xi^{\pm}\rangle$. Consequently, we retain only the results corresponding to the measurement outcome of $|\eta^{\pm}\rangle$ and discard those corresponding to the measurement outcome $|\xi^{\pm}\rangle$. This selective approach ensures that the fidelity of the final entangled state is optimized, thus improving the overall efficiency of the entanglement swapping process.

To employ measurement on Charlie's qubits, the measurement operators are defined as
\begin{equation}
    C_{\eta\pm}=I\otimes I \otimes |\eta^{\pm}\rangle \langle\eta^{\pm}| \otimes I,
\end{equation}
where $I$ is the identity operator and $|\eta^{\pm}\rangle$ are defined in Eq. \eqref{orth}.

The normalized final shared entangled state between Alice and Bob, corresponding to the measurement outcome $|{\eta ^{\pm}}\rangle$ and taking into account the effects of ADC, can be represented as 
\begin{equation}\label{rho_adc}
    \begin{aligned}
        \rho_{126}^{\text{AD}} =\text{Tr}_{345}\left( C_{\eta\pm} \ {\
rho}^{\text{AD}}_{\text{combined}} \ C_{\eta\pm}^{\dagger}\right)=|{W}\rangle_{126}^{\text{AD}} \langle{W}|_{126}^{\text{AD}} + \frac{r}{1+r}  |{000}\rangle\langle{000}|,
    \end{aligned}
\end{equation}
where $|{W}\rangle_{126}^{\text{AD}} = \frac{1}{2\sqrt{1+r}} \left(  |{100}\rangle+  |{010}\rangle + \sqrt{2} |{001}\rangle \right)$. 

The probability of obtaining the measurement outcome $|{\eta ^{\pm}}\rangle$ is given as
\begin{equation}\label{g-eta}
 g_{\eta\pm} = \text{Trace}\left( \text{Tr}_{345}\left( C_{\eta\pm} \ {\rho}^{\text{AD}}_{\text{combined}} \ C_{\eta\pm}^{\dagger}\right)\right)=\frac{1-r^2}{4}.
\end{equation}

The fidelity of the final shared entangled state, considering the effects of the amplitude damping is derived as 
\begin{equation}\label{fid_er}
    \text{Fid}^{\text{AD}}=\langle W\lvert\rho^{\text{AD}}_{126}\rvert W\rangle=\frac{1}{1+r}.
\end{equation}

Now, we provide the details of our weak measurement-based purification method to improve the fidelity of the final shared entangled state in presence of amplitude damping. 
Following the entanglement swapping process, Alice and Bob must perform weak measurements on their respective qubits of the damped shared entangled state, as represented in Eq. \eqref{rho_adc}. The weak measurement operator is from the complete measurement set \(\{M,\overline{M}\}\), respectively, as \cite{lab12g}:
\begin{equation}\label{WM-op}
     M = \begin{pmatrix}
   \sqrt{1 - q} & 0 \\
   0 & 1
   \end{pmatrix}, \quad
   \overline{M} = \begin{pmatrix}
   \sqrt{q} & 0 \\
   0 & 0
   \end{pmatrix},
\end{equation}
where ${M}^\dagger{M}+{\overline{M}}^\dagger\overline{M}=I$ and $q\in[0,1]$  represents the strength of the weak measurement. In our scheme, we selectively keep the result of $M$ while discarding the outcome of $\overline{M}$ to improve the fidelity of the final shared entangled state. 

The weak measurements employed in our scheme have been experimentally implemented in photonic architectures \cite{cite1, cite2} and nuclear magnetic resonance systems \cite{cite2}, demonstrating its feasibility and practical relevance.

Both Alice and Bob are required to conduct weak measurements on their respective qubits. The normalized final shared entangled state following the application of weak measurements, based on the joint measurement outcomes $|\eta^{\pm}\rangle$, can be calculated as
\begin{equation}\label{rho_wm}
    \begin{aligned}
        \rho_{126}^{\text{WM}} &= (M \otimes M \otimes M)\rho_{126}^{\text{AD}}(M \otimes M \otimes M)^{{\dagger}}\\
        &=\frac{1}{\text{p}^{\text{WM}}}\times\frac{(1-q)^2}{1+r}\Bigg(|{W}\rangle_{126}\langle{W}|_{126} + (r(1-q))  |{000}\rangle\langle{000}|\bigg),
    \end{aligned}
\end{equation} 
where $|{W}\rangle_{126}=|W\rangle $ given in Eq. \eqref{Eqw} and $\text{p}^{\text{WM}}$ is the success probability of employing weak measurements given as
\begin{equation}    
\begin{aligned}
\label{P-WM}
    \text{p}^{\text{WM}}&=\text{trace}((M \otimes M \otimes M)\rho_{126}^{\text{AD}}(M \otimes M \otimes M)^{\dagger})=\frac{(1-q)^2(r-qr+1)}{1+r}.  
\end{aligned}
\end{equation}

The improved fidelity achieved through the application of weak measurements can be expressed as
\begin{equation}\label{fid_wm}
    \text{Fid}^{\text{WM}}=\langle W\lvert\rho_{126}^{\text{WM}}\rvert W\rangle=\frac{1}{1+r(1-q)}.
\end{equation}

\begin{figure*}[htbp]
    \centering
    \hspace*{-10pt}
    \subfigure[]{
        \centering
        \includegraphics[width=0.45\textwidth]{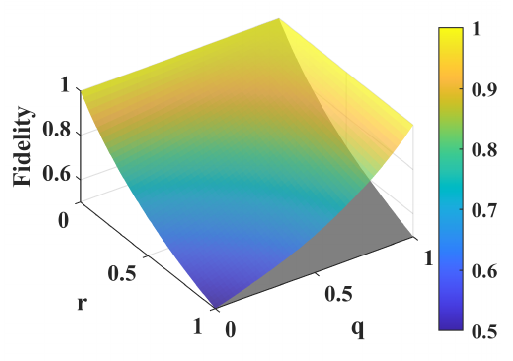}
        \label{Fig2a}
    }%
    \subfigure[]{
        \centering
        \includegraphics[width=0.45\textwidth]{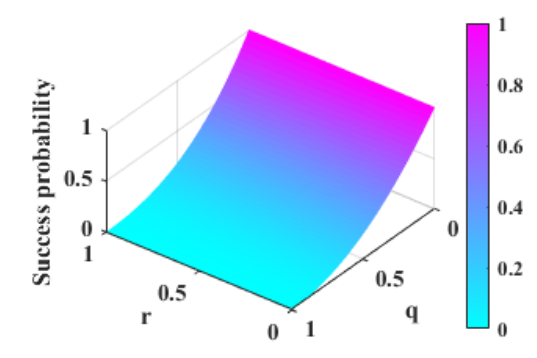}
        \label{Fig2b}
    }%
\caption{(a) Fidelity of the final shared entangled state as a function of decay rate $r$ and weak measurement strength $q$. The colored surface represents the fidelity of the shared entangled state after employing weak measurements, while the gray surface represents the fidelity without any purification. (b) Success probability of weak measurements as a function of decay rate $r$ and weak measurement strength $q$.}
\label{Fig2}
\end{figure*}

Fig. \ref{Fig2a} depicts the fidelity of the final shared entangled state between Alice and Bob for the measurement outcome $|{\eta ^{\pm}}\rangle$ as a function of the decay rate $r$ and weak measurement strength $q$. The colored surface represents the fidelity after employing weak measurements as described in Eq. \eqref{fid_wm}, while the gray surface corresponds to the fidelity without any purification process, as given in Eq. \eqref{fid_er}. Fig. \ref{Fig2b} demonstrates the success probability of weak measurements as a function of the decay rate $r$ and weak measurement strength $q$.

As shown in Fig. \ref{Fig2}, the fidelity of the shared entangled state is significantly enhanced through the use of weak measurements. However, there is a trade-off between fidelity and success probability, as higher values of $q$ result in increased fidelity but lower success probabilities associated with weak measurements, as demonstrated in Fig. \ref{Fig2b}. Therefore, it is crucial to strike a balance between fidelity and success probability based on the specific requirements and priorities of the given situation. This can be achieved by either opting for larger $q$ values for higher fidelity with lower success probability or smaller $q$ values for lower fidelity with higher success probability.

In order to determine the total probability of success for the entanglement swapping process, it is crucial to take into account two factors: the probability of obtaining the measurement outcome $|\eta^{\pm}\rangle$, $g_{\eta\pm}$ in Eq. \eqref{g-eta} and the success probability associated with the weak measurements, $ p^{\text{WM}}$ in Eq. \eqref{P-WM}. Hence, the total success probability of the entanglement swapping process is defined as
\begin{equation}
\begin{aligned}
   P^{\text{WM}}_{\text{swapping}}&=2 \times \text{p}^{\text{WM}} \times g_{\eta\pm}=\frac{(1-q^2)(1-r)(r-qr+1)}{2}.
\end{aligned}
\end{equation}

\begin{figure}[!htpb]
    \centering
    \includegraphics[width=0.4\linewidth]{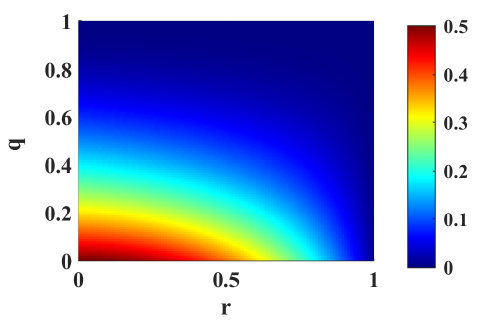}
    \caption{Entanglement swapping success probability by employing weak measurement-based purification method.}
    \label{swap_p}
\end{figure}

Fig. \ref{swap_p} displays the success probability of entanglement swapping by considering weak measurement-based purification methods as a function of $q$ and $r$.

As illustrated in Fig. \ref{swap_p}, a smaller value of $q$ leads to a higher overall success probability for the combined process of swapping and purification. It is important to note that, since we discard the results corresponding to the measurement outcome $|\xi^{\pm}\rangle$, the maximum achievable success probability is 0.5.

\subsection{Qiskit implementation of weak measurement-based purification method}

\begin{figure}[!htpb]
    \centering
    \includegraphics[width=\linewidth]{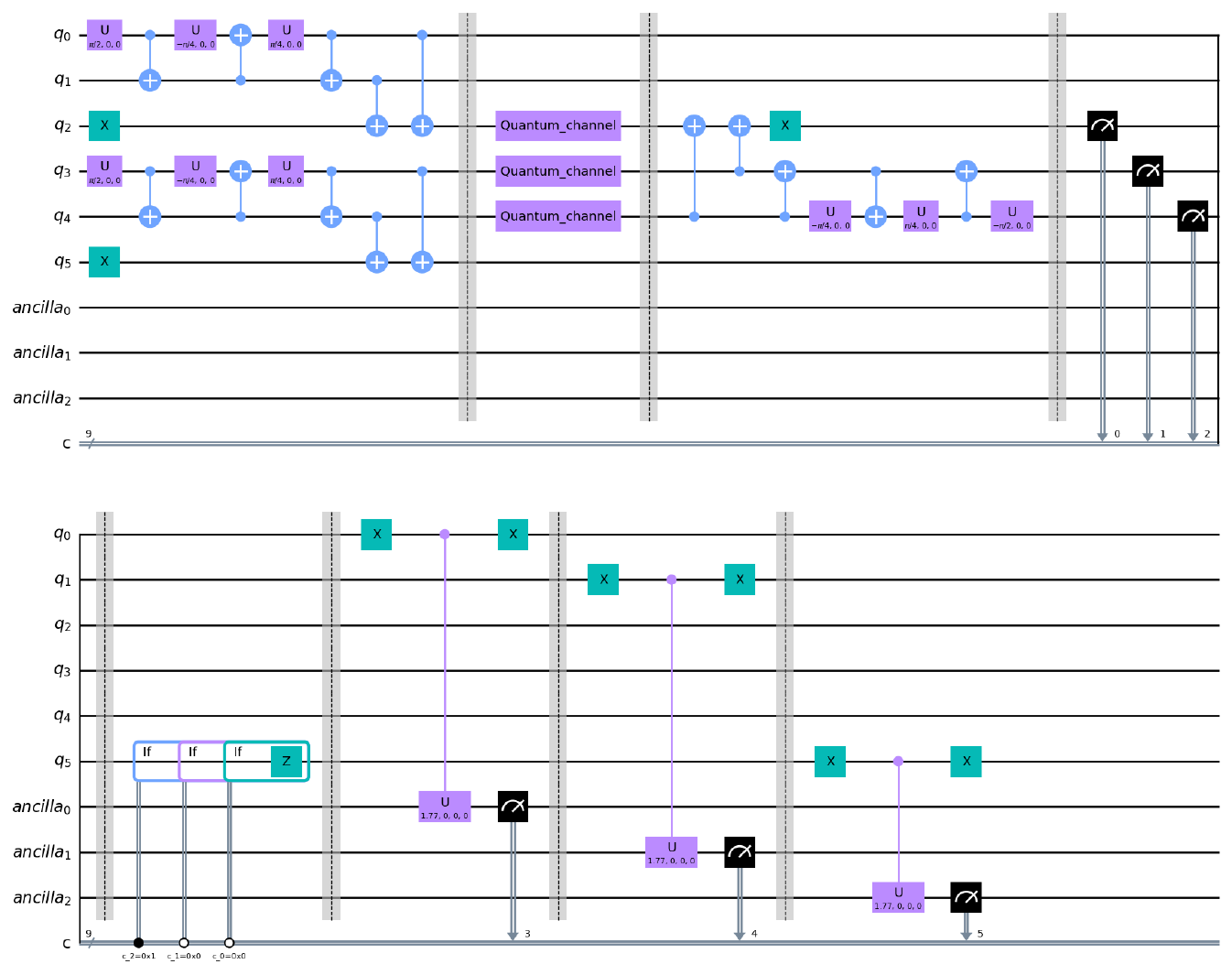}
    \caption{Qiskit circuit of the entanglement swapping of the W state through amplitude damping channels and the application of our proposed weak measurement-based purification method.}
    \label{wm-qiskit}
\end{figure}

In this subsection, we demonstrate the practical implementation of our proposed weak measurement-based purification method using Qiskit. The designed Qiskit circuit is presented in Fig. \ref{wm-qiskit}.

Prior to the first barrier, the specific  W states are prepared. Subsequently, the third qubit of the first W state and the first two qubits of the second W state are sent through the ADCs. Following the second barrier, the designed joint measurements are performed. After the third barrier, the qubits at Charlie's location are measured, and corresponding unilateral operations are applied to the last qubit of the shared W state after the fourth barrier, based on the measurement results. Since we only consider the outcomes corresponding to the measurement results $|{\eta ^{\pm}}\rangle$, we apply a $z$-rotation only if $|{\eta ^{-}}\rangle$ is obtained. After the fifth barrier, the implementation of the weak measurements takes place, completing the purification process for the entangled W state.

To implement the weak measurements, we need to add ancilla qubits to the circuit and use a controlled rotation gate configuring the W state qubit as the control qubit and the ancilla qubit as the target qubit. The specific mathematical formula of the Qiskit's controlled rotation gate is given as
\begin{widetext}
\begin{equation}
    C_u(\theta,\phi,\lambda, \gamma)=\begin{bmatrix}
        1&0&0&0\\0&e^{-i(\phi+\lambda)/2}\cos(\theta/2)&0&-e^{-i(\phi-\lambda)/2}\sin(\theta/2)\\
       0&0&1&0\\0&e^{i(\phi-\lambda)/2}\sin(\theta/2)&0&e^{i(\phi+\lambda)/2}\cos(\theta/2)
    \end{bmatrix}.
\end{equation}
  
\end{widetext}

In order to implement the weak measurement in Eq. \ref{WM-op}, the controlled rotation gate parameters should be set as $\theta=2\arctan(\frac{\sqrt{q}}{\sqrt{1-q}}) \text{ and } \phi=\lambda=\gamma=0 $, where $q$ is the strength of the weak measurement. Additionally, we need to apply an $x$-rotation gate before the controlled rotation gate to achieve the same effects as the weak measurement operators. Subsequently, another $x$-rotation gate is employed after the controlled rotation gate to restore the W state to its original structure.

\begin{figure}[!htpb]
    \centering
    \includegraphics[width=0.4\linewidth]{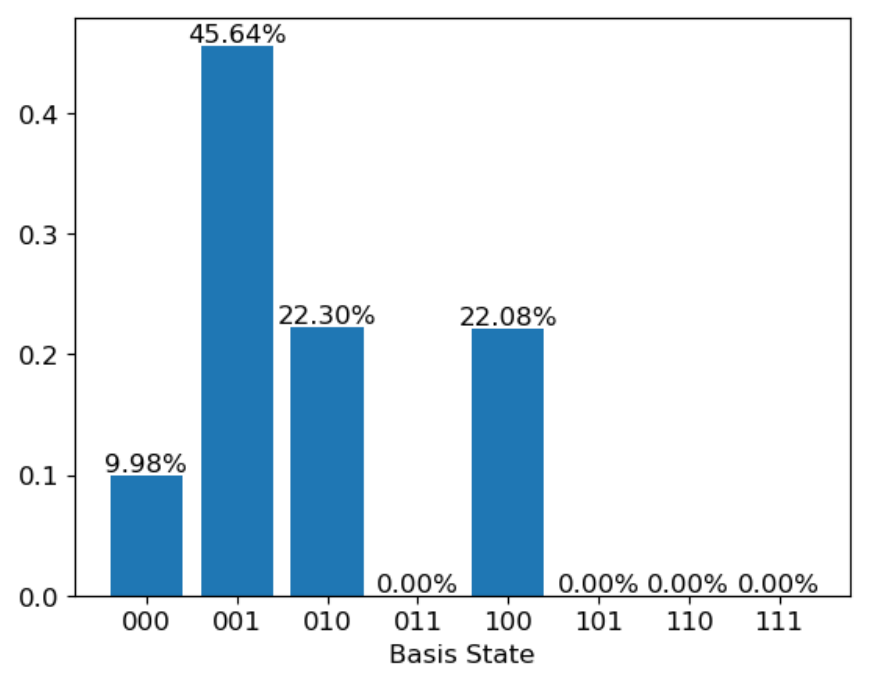}
    \caption{Average probabilities of basis states of the final shared entangled state between Alice and Bob after employing weak measurement-based purification method. Here $r=0.3$ and $q=0.64$.}
    \label{wm-state}
\end{figure}

By executing the Qiskit circuit on the AerSimulator and subsequently tracing out the ancilla qubits, we selectively consider states corresponding to ancilla qubit measurement outcomes of 0. Furthermore, we select the states corresponding to Charlie's measurement outcomes $|{\eta ^{\pm}}\rangle$. The resulting average probabilities of basis states are depicted in Fig. \ref{wm-state}.

By setting the same parameters as $r=0.3$ and $q=0.64$, it is evident that the final state aligns precisely with Eq. \eqref{rho_wm}. This observation substantiates the accuracy and validity of our mathematical derivations through Qiskit simulation.

\section{Conclusion}\label{Sec3}

In this paper, we have introduced a deterministic entanglement swapping protocol custom-designed for the generation of W states – a crucial resource for quantum information processing tasks. Our approach offers a more dependable and efficient solution for quantum communication networks, surpassing the performance of probabilistic methods. We have provided a practical framework for real-world applications by outlining a Qiskit-based quantum circuit that facilitates the preparation of W states and the execution of joint measurements required for the entanglement swapping process. Furthermore, we have examined the impact of imperfect operations and noisy channels on the fidelity of the shared W state and proposed a weak measurement-based purification method to enhance fidelity in the presence of amplitude damping. Both mathematical analysis and Qiskit simulations have been employed to validate the reliability and relevance of our proposed entanglement swapping protocol, demonstrating its effectiveness in practical quantum communication scenarios. Our findings contribute to the advancement of quantum communication technologies, paving the way for more robust and efficient quantum networks in the future.

\section*{Acknowledgement}
This work was supported by the National Natural Science Foundation of China under Grant Number 61973290.

\bibliographystyle{quantum}
\bibliography{reference}

\end{document}